\documentclass{sig-alternate}
\usepackage{times,epsfig, graphicx, subfig, float}
\usepackage{url,endnotes}
\usepackage{algorithm}
\usepackage{algorithmic}
\usepackage{microtype}
\usepackage{times,setspace}
\usepackage{latexsym}
\usepackage{amsmath}
\usepackage{url}
\usepackage{mdwlist}
\usepackage{pifont}
\usepackage{fixltx2e}
\usepackage{flushend}
\usepackage{pifont}
\usepackage[usenames,table]{xcolor}
\DeclareCaptionType{copyrightbox}

\begin{document}

\conferenceinfo{ASIA CCS'13,} {May 8--10, 2013, Hangzhou, China.} 
\CopyrightYear{2013} 
\crdata{978-1-4503-1767-2/13/05} 
\clubpenalty=10000 
\widowpenalty = 10000

\title{SecLaaS: Secure Logging-as-a-Service for Cloud Forensics}
\numberofauthors{3} 
\author{
\alignauthor
Shams Zawoad\\
		\affaddr{University of Alabama at Birmingham}\\
		\affaddr{Birmingham, Alabama 35294-1170}\\
       \email{zawoad@cis.uab.edu}
\alignauthor
Amit Kumar Dutta\\
		\affaddr{University of Alabama at Birmingham}\\
		\affaddr{Birmingham, Alabama 35294-1170}\\	
       \email{adutta@cis.uab.edu}
\alignauthor
Ragib Hasan\\
		\affaddr{University of Alabama at Birmingham}\\
		\affaddr{Birmingham, Alabama 35294-1170}\\	
       \email{ragib@cis.uab.edu}       
}

\maketitle

\newpage
\begin{abstract}
Cloud computing has emerged as a popular computing paradigm in recent years. However, today's cloud computing architectures often lack support for computer forensic investigations. Analyzing various logs (e.g., process logs, network logs) plays a vital role in computer forensics. Unfortunately, collecting logs from a cloud is very hard given the black-box nature of clouds and the multi-tenant cloud models, where many users share the same processing and network resources. Researchers have proposed using log API or cloud management console to mitigate the challenges of collecting logs from cloud infrastructure. However, there has been no concrete work, which shows how to provide cloud logs to investigator while preserving users' privacy and integrity of the logs. In this paper, we introduce Secure-Logging-as-a-Service (\textit{SecLaaS}), which stores virtual machines' logs and provides access to forensic investigators ensuring the confidentiality of the cloud users. Additionally, \textit{SeclaaS} preserves proofs of past log and thus   protects the integrity of the logs from dishonest investigators or cloud providers. Finally, we evaluate the feasibility of the scheme by implementing \textit{SecLaaS} for network access logs in OpenStack -- a popular open source cloud platform. 
\end{abstract}




\section{Introduction}
\label{sec:introduction}

Cloud computing offers infinite infrastructure resources, very convenient pay-as-you-go service, and low cost computing. As a result, cloud computing has become one of the most dominant computing paradigms in recent years. Today, small and high level companies are attracted to cloud computing because it does not require any kind of local infrastructure setup, and, at the same time, it is highly cost effective. According to Khajeh-hossainei, an organization can save 37\% of its cost just by moving their IT infrastructure from an outsourced data center to Amazon's cloud \cite{khajeh2010cloud}. Market Research Media stated in one of their recent reports that the cloud computing market is expected to grow at a 30\% compound annual growth rate and will reach \$270 billion in 2020 \cite{websitemarketresearchmedia}. Garner Inc. states that the strong growth of cloud computing will bring \$148.8 billion revenue by 2014 \cite{websitegartnernews}. From the research work of INPUT, it is clear that Cloud computing is equally popular in both Government and private industry; their report identifies that the federal cloud market is expected to expand to \$800 million by 2013 \cite{input2009}.

Cloud computing opens a new horizon of computing for business and IT organizations. However, at the same time, malicious individuals can easily exploit the power of cloud computing. Attackers can attack applications running inside the cloud. Alternatively, they can launch attacks from machines inside the cloud. These issues are the primary concerns of \textit{Cloud Forensics}. An annual report of the Federal Bureau of Investigation (FBI) states that, the size of the average digital forensic case is growing 35\% per year in the United States. From 2003 to 2007, it increased from 83GB to 277 GB \cite{fbi2008}. As a result, forensic experts are devising new techniques for digital forensics. There are several forensics analysis schemes and tools available in market. Unfortunately, none of them are suitable for the dynamic nature of cloud computing. Many of the implicit assumptions made in regular forensics analysis (e.g., physical access to hardware) are not valid for cloud computing. Hence, for cloud infrastructure, a special branch of digital forensics has been brought up by researchers - \textit{Cloud Forensics}. Cloud forensics offers new challenges and has opened new research problems for security and forensics experts, which are important from both technical and legal point of view.
	
The process of digital forensics starts with acquiring the digital evidence. In a cloud, the evidence could be the image of virtual machines, files stored in cloud storage, and logs provided by cloud service providers (CSP). However, collecting these evidences, specially logs from cloud infrastructure, is extremely difficult because cloud users or investigators have very little control over the infrastructure. Currently, to collect logs from cloud, investigators are dependent on the CSP. Investigators need to issue a subpoena to the CSP to acquire the logs of a particular user. However, they need to believe the CSPs blindly, as there is no way to verify whether the CSPs are providing valid logs or not. Moreover, if an adversary shuts down the virtual machine (VM) she is using, there is no way to collect logs from the terminated VM. 

To overcome the challenges of acquiring logs from cloud infrastructure, Bark et al. proposed that the CSPs can provide network, process and access logs to customer by a read-only API \cite{birk2011technicalIssues}. To solve the same problem, Dykstra et al. recommended a cloud management plane for using in Infrastructure-as-a-Service model \cite{dykstraacquiring}. However, they did not show how we can practically implement those schemes. Additionally, log information is highly sensitive and user's privacy issues are directly related to it. Previous studies do not provide a secure way of revealing the logs while maintaining user privacy. Moreover, it is vital to ensure that logs are not tampered with before exposing to investigators. For a successful forensic scheme based on logs, these issues must be resolved in a secure and trustworthy manner. 

In this paper, we take the first step towards exposing a publicly available secure log service. This service can be used by forensic investigators to identify malicious activities that took place in virtual machines of a cloud system.

To illustrate the specific problem we look at, we present the following hypothetical scenario:\smallskip

\emph{Alice is a successful businesswoman who runs a shopping website in cloud. The site serves a number of customers every day and her organization generates a significant amount of profit from it. Therefore, if the site is down even for a few minutes, it will seriously hamper not only their profit but also the goodwill. Mallory, a malicious attacker decided to attack Alice's shopping website. She rented some machines in cloud and launched a Distributed Denial of Service (DDoS) attack to the shopping website using those rented machines. As a result, the site was down for an hour, which had quite negative impact on Alice's business. Consequently, Alice asked a forensic investigator to investigate the case. The investigator found that Alice's website records the visiting customer's IP address. Analyzing the visiting customers records, the investigator found that Alice's website was flooded by some IP addresses which are owned by a cloud service provider. Eventually, the investigator issued a subpoena to the corresponding cloud provider to provide him the network logs for those particular IP addresses. On the other hand, Mallory managed to collude with the cloud provider after the attack. Therefore, while providing the logs to the investigator, the cloud provider supplied tampered log to the investigator, who had no way to verify the correctness of the logs. Under this circumstance, Mallory will remain undetected. Even if the cloud provider was honest, Mallory could terminate her rented machines and left no traces of the attack. Hence, the cloud provider could not give any useful logs to the investigator.}\smallskip 	

To mitigate the challenges discussed in the above scenario, we propose the notion of Secure-Logging-as-a-Service (SecLaaS) in this paper. \\

\noindent\textbf{Contributions:~} The contributions of this paper are as follows:
\begin{enumerate}
\item We propose a scheme of revealing cloud users' logs for forensics investigation while preserving the confidentiality of users' logs from malicious cloud employee or external entity;
\item We introduce Proof of Past Log (PPL) -- a tamper evident scheme to prevent the cloud service provider or investigators from manipulating the logs after-the-fact.
\item We evaluate the proposed scheme using a open source cloud computing platform.\\
\end{enumerate}
	
\noindent\textbf{Organization:~} The rest of this paper is organized as follows. Section~\ref{sec:background} provides some background information and challenges of cloud forensics in terms of logging. Section~\ref{sec:threatmodel} describes the adversary's capabilities and possible attacks on logging-as-a-service. Section~\ref{sec:scheme} presents our SecLaaS scheme and Section \ref{sec:security} provides security analysis of the scheme. In Section \ref{sec:implementation}, we provide the implementation and performance evaluation of our scheme on an open source cloud software, OpenStack. Section~\ref{sec:discussion} discusses the usability of our proposed schemes. In Section~\ref{sec:relatedwork}, we provide an overview of related research about logging in cloud forensics, and finally, we conclude in Section~\ref{sec:conclusion}.

\section{Background and Challenges}
\label{sec:background}
With the increasing popularity of cloud computing, there is a significant interest in the law-enforcement community to extend digital forensics techniques in the context of a cloud. In this section, we present the definitions of digital forensics and cloud forensics, motivation behind our work, and discuss the challenges of logging-as-a-service for cloud forensics.

\subsection{Digital Forensics}
Digital forensics is the process of preserving, collecting, confirming, identifying, analyzing, recording, and presenting crime scene information. Wolfe defines digital forensics as \emph{``A methodical series of techniques and procedures for gathering evidence, from computing equipment and various storage devices and digital media, that can be presented in a court of law in a coherent and meaningful format"} \cite{wiles2007best}.  According to a definition by NIST \cite{kent2006guide}, computer forensics is an applied science to identify an incident, collection, examination, and analysis of evidence data. While doing so, maintaining the integrity of the information and strict chain of custody for the data is mandatory. Several other researchers define computer forensics as the procedure of examining computer system to determine potential legal evidence \cite{lunn2000computer,robbins2008explanation}. 

\subsection{Cloud Forensics}
Cloud forensics can be defined as applying computer forensics procedures in a cloud computing environment. As cloud computing is based on extensive network access, and as network forensics handles forensic investigation in private and public network, Ruan et al. defined cloud forensics as a subset of network forensics \cite{ruan2011cloud}. They also identified three dimensions in cloud forensics -- technical, organizational, and legal. Cloud forensics procedures will vary according to the service and deployment model of cloud computing. For Software-as-a-Service (SaaS) and Platform-as-a-Service (PaaS), we have very limited control over process or network monitoring. Whereas, we can gain more control in Infrastructure-as-a-Service (IaaS) and can deploy some forensic friendly logging mechanism. The first three steps of computer forensics, identification, collection, and organization of evidence will vary for different service and deployment model. For example, the evidence collection procedure of SaaS and IaaS will not be same.  For SaaS, we solely depend on the CSP to get the application log, while in IaaS, we can acquire the virtual machine image from the customer and can enter into examination and analysis phase. On the other hand, in the private deployment model, we have physical access to the digital evidence, but we merely can get physical access to public deployment model.

\subsection{Motivation}

Though cloud computing offers numerous opportunities to different level of consumers, many security issues of cloud environment have not been resolved yet. According to a recent IDCI survey, 74\% of IT executives and CIO's referred to security as the main reason to prevent their migration to the cloud services model \cite{websiteclavister}. Some recent and well-publicized attacks on cloud computing platform justify the concern with security. For example, a botnet attack on Amazon's cloud infrastructure was reported in 2009 \cite{websiteamazon2009}. 

Besides attacking cloud infrastructure, adversaries can use the cloud  to launch attack on other systems. For example, an adversary can rent hundreds of virtual machines to launch a Distributed Denial of Service (DDoS) attack. After a successful attack, she can erase all the traces of the attack by turning off the virtual machines. A criminal can also keep her secret files (e.g., child pornography, terrorist documents) in cloud storage and can destroy all her local evidence to remain clean. When law enforcement investigates such a suspect, the suspect can deny launching a DDoS attack. At present, there is no way to claim that an adversary access a certain network at a given time. 

Researchers are working to protect the cloud environment from different types of attacks. However, in case of an attack, we also need to investigate the incident, i.e., we need to carry out a digital forensic investigation in the cloud. Besides protecting the cloud, we need to focus on this issue. Unfortunately, there has been little research on adapting digital forensics for use in cloud environments. In this paper, we address this problem, which has significant real-life implications in law enforcement investigating cybercrime and terrorism.

\subsection{Challenges}
Analyzing logs from different processes plays a vital role in digital forensic investigation. Process logs, network logs, and application logs are really useful to identify a malicious user. However, gathering this crucial information in cloud environment is not as simple as it is in privately owned computer system, sometimes even impossible. The inherent characteristics of cloud have made the forensic log-analysis a nightmare for the forensic investigators. It is very difficult to collect and prove the validity of the logs to the court authority. For example, how can an investigator collect network logs of malicious VMs, which have been already terminated by the attacker after launching a DDoS attack last month? We must find secure techniques for storing and providing logs to investigators, which also need to be admissible in a court of law as valid evidence. Many things can complicate the log collection process. A malicious CSP can change the logs while providing the logs to investigators. Clients may question the integrity of any such logs, claiming that the forensic investigators or the prosecution and the CSP have colluded to plant evidence in the cloud. The following reasons also make the log collection and providing the proof of the logs challenging in cloud. \\

\begin{figure}[t]
\centering
\includegraphics[width=0.48\textwidth]{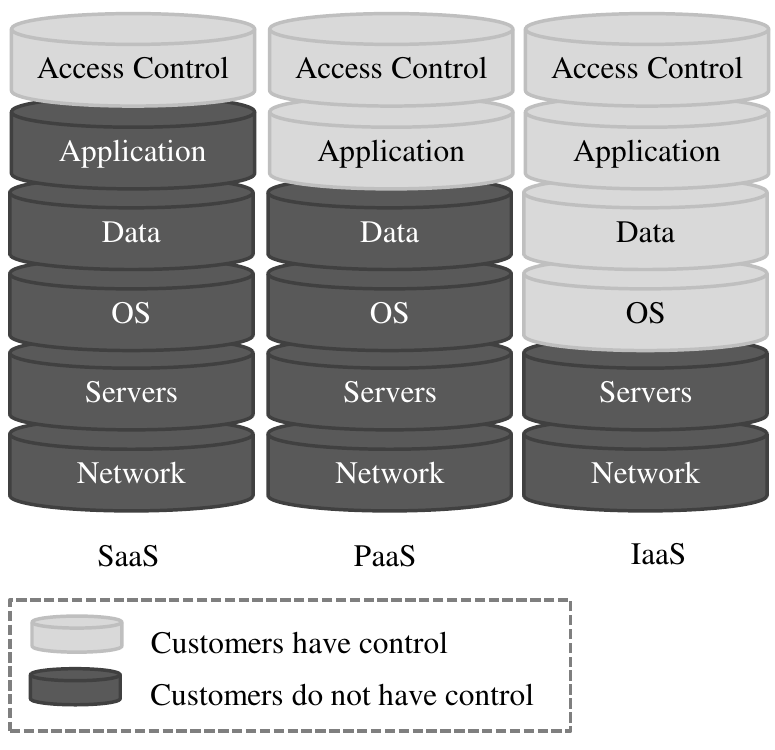}
\caption{Customers' control over different layers in different service model}
\label{figure:servicecontrol}
\end{figure}

\noindent\textbf{Reduced Level of Control, and Dependence on the CSP:~} One of the challenges of collecting logs securely from cloud is the users' or investigators' reduced level of control over the cloud environment. In traditional computer forensics, the investigators have full control over the evidence (e.g., router logs, process logs, hard disk). Currently, to acquire the logs, we extensively depend on the CSPs. The availability of the logs varies depending on the service model. Figure \ref{figure:servicecontrol} shows the control of customers in different layers for the three different service models -- IaaS, PaaS, and SaaS. From the figure, we can observe that cloud users have highest control in IaaS and least control in SaaS. This physical inaccessibility of the evidence and lack of control over the system make evidence acquisition a challenging task in cloud forensics. In SaaS, customers do no get any log of their system, unless the CSP provides the logs. In PaaS, it is only possible to get the application log from the customers. To get the network log, database log, or operating system log we need to depend on the CSP. For example, Amazon does not provide load balancer log to the customers \cite{websiteamazonwebservice}. In a recent research work, Marty mentioned that he was unable to get MySql log data from Amazon's Relational Database Service \cite{marty2011cloud}. In IaaS, customers do not have the network or process logs. Several other problems come along with the less control issue. For example, we need to depend on the cloud service providers for evidence acquisition, which in turn brings the honesty issue of the CSP's employee, who is not a certified forensic investigator. CSPs can always tamper the logs as they have the full control over the generated logs. Additionally, CSPs are not always obligated to provide all the necessary logs.\\

\noindent \textbf{Decentralization.~} In cloud infrastructure, log information is not located at any single centralized log server; rather logs are decentralized among several servers. Multiple users' log information may be co-located or spread across multiple servers. Moreover, there are several layers and tiers in cloud architecture. Logs are generated in each tier. For example, application, network, operating system, and database -- all of these layers produce valuable logs for forensic investigation. Collecting logs from these multiple servers and layers and providing it to investigators in a secure way is extremely challenging.\\

\noindent \textbf{Accessibility of Logs.~} The logs generated in different layers are required to be accessible to different stakeholders of the system, e.g., system administrator, forensic investigator, and developer. System administrators need relevant log to troubleshoot the system; developers need the required log to fix the bug of the application; forensic investigators need logs, which can help in their investigation. Hence, there should be some access control mechanism, so that everybody will get what they need exactly -- nothing more, nothing less and obviously, in a secure way. We should not expect that a malicious cloud employee, who can violate the privacy of the users gets access to users' log information.\\

\noindent\textbf{Multi-tenancy:~} In cloud computing, multiple virtual machines (VM) can share the same physical infrastructure, i.e., log for multiple customers may be co-located. The nature of this infrastructure is different from the traditional single owner computer system. Hence, while collecting logs for one user, other users' data can be mingled with the log evidence. An alleged user can claim that the log contains information of other users, not her. The investigator then needs to prove it to the court that the provided logs indeed belongs to the malicious user. Moreover, we need to preserve the privacy of the other tenants. For both of these issues, collecting and providing logs to the investigator is challenging in cloud paradigm.\\

\noindent\textbf{Chain of custody:~} The chain of custody is one of the most vital issues in traditional digital forensic investigation. The chain of custody should clearly depict how the evidence was collected, analyzed, and preserved in order to be presented as admissible evidence in court \cite{vacca2005computer}. In traditional forensic procedure, it starts with gaining the physical control of the evidence, e.g., computer, hard disk. However, in cloud forensics, this step is not possible due to the multi jurisdictional laws, procedures, and proprietary technology in cloud environment \cite{taylor2010digital,grisposcalm}. Collecting logs from cloud infrastructure, analyzing, and presenting the proof of logs need to follow this chain of custody. We must clarify certain things to maintain the chain of custody, e.g., how the logs were generated and stored, and who had the access to the logs.\\

\noindent\textbf{Presentation:~} The final step of digital forensic investigation is presentation, where an investigator accumulates his findings and presents to the court as the evidence of a case. Challenges also lie in this step of cloud forensics \cite{reilly2011cloud}. Proving the integrity of network and process logs in front of jury for traditional computer forensics is relatively easy, compared to the complex structure of cloud computing. Presenting the logs and the proofs of integrity of the logs to the court in an admissible way is challenging for cloud computing. 
\section{Threat Model}
\label{sec:threatmodel}

In this section, we describe the attacker's capability, possible attacks on logs, and properties of our proposed system. Before describing the threat model, we first define the important terms to clarify the threat model.
\subsection{Definition of terms}
\begin{itemize}

\item \emph{User:} A user is a customer of the cloud service provider (CSP), who uses the CSP's storage service. A user can be malicious or honest.

\item \emph{Log:} A log can be the network log, process log, operating system logs, or any other logs generated in cloud for a VM.

\item \emph{Proof of Past Logs (PPL):} The \textit{PPL} contains the proof of logs to verify whether some logs belong to a particular user or not.

\item \emph{Investigator:}  An investigator is a professional forensic investigator, who needs to collect necessary logs from cloud infrastructure in case of any malicious incident.

\item \emph{CSP:} The Cloud Service Provider (CSP) will generate the \textit{PPL} and give access to the logs to users and investigators through an API or management console.

\item \emph{Log Chain (LC):} The \textit{LC} maintains the correct order of the logs. From the \textit{LC}, it can be verified that the CSP or the investigators provide logs in the actual order of log generation.

\item \emph{Auditor:} Most likely, the auditor will be the court authority that will verify the correctness of the logs from the \textit{PPL} and \textit{LC}.

\item \emph{Intruder:} An intruder can be any malicious person including a employee of the CSP, who wants to reveal user's activity from the \textit{PPL} or from the log storage. 

\end{itemize}

\subsection{Attacker's Capability}
In our threat model, we assume that the users and the investigators do not trust the CSPs, and both of them can be malicious. We assume that a CSP is honest at the time of publishing the \textit{PPL} and \textit{LC}. However, during the investigation, CSP can collude with a user or an investigator and provide tampered logs, for which the \textit{PPL} and \textit{LC} have already been published. User, investigator, and CSP can collude with each other to provide fake logs to the auditor. A user cannot modify the logs by him, but he can collude with CSP to alter the logs. An investigator can present false log information to the court to frame an honest user  or can collude with a malicious user to save her from accusation.  The CSP can also repudiate any published \textit{PPL}. An intruder can acquire the \textit{PPL} of a user to learn the user's activity. A malicious cloud employee can also be an intruder. 

\subsection{Possible Attacks}
There can be different types of attacks on providing log API. CSP can remove some crucial logs or can reorder the logs. A user can deny the ownership of any logs. Even an investigator can present invalid logs to the court. Below we mention some of the possible attacks:

\begin{itemize}
\item \emph{Privacy violation:} If the CSP published the PPL publicly on the web, any malicious person can acquire the published PPL and try to learn about the logs from the proof. Even if logs are kept unpublished, an otherwise honest employee of the CSP who has access to the log storage can identify the activity of the user from the stored logs.

\item \emph{Log modification:} A dishonest CSP, while colluding with user or investigator can modify the logs, either to save a malicious user or to frame a honest user. If an investigator is not trustworthy, he can also tamper with the logs before presenting the logs to the court. There can be three types of contamination of logs:
	\begin{enumerate}
		\item Removal of crucial logs
		\item Planting of false logs
		\item Modification of the order of the logs 
	\end{enumerate}	 
 
\item \emph{Repudiation by CSP:} An otherwise honest CSP can deny a published PPL/LC after-the-fact.

\item \emph{Repudiation by User:} As data are co-mingled in the cloud, a malicious user can claim that the logs contain another cloud user's data.
 
\end{itemize}

\subsection{System Property}
In designing SecLaaS, our goal is to ensure the secure preservation of cloud users' logs in a persistent storage. Our mechanism should prevent any malicious party to produce a false proof of past logs \textit{PPL}. A false \textit{PPL} attests the presence of a log record for a user, which the user does not actually own. Once the proof has been published, the CSP can neither modify the proof nor repudiate any published proof. Additionally, we must prevent false implications by malicious forensic investigators. Based on our analysis, a secure log service for clouds should possess the following integrity and confidentiality properties:

\begin{itemize}
 \item\textbf{I1:} The CSP cannot remove a log entry from the storage after publishing the PPL.

 \item\textbf{I2:} The CSP cannot change the order of a log from its actual order of generation.	

 \item\textbf{I3:} The CSP cannot plant false log after-the-fact.
 
 \item\textbf{I4:} An investigator cannot hide or remove a log entry at the time of presenting logs to court.
 
 \item\textbf{I5:} An investigator cannot change the actual order of a log entry at the time of presenting evidences to court.
 
 \item\textbf{I6:} An Investigator cannot present phony log to the court.

 \item\textbf{I7:} The CSP cannot repudiate any previously published proof of logs.

 \item\textbf{C1:} From the published proof of log, no adversaries can recover any log.

 \item\textbf{C2:} A malicious cloud employee will not be able to recover logs from the log storage.

\end{itemize}
\section{The S\lowercase{ec}L\lowercase{aa}S Scheme}
\label{sec:scheme}
In this section, we present SecLaaS -- our system for secure retrieval of logs and storage of the proof of past logs. Initially, we provide an overview of the mechanism, followed by the schematic and the protocol specific description of the system.

\subsection{Overview}
\begin{figure}[!ht]
\centering
\includegraphics[width=0.48\textwidth]{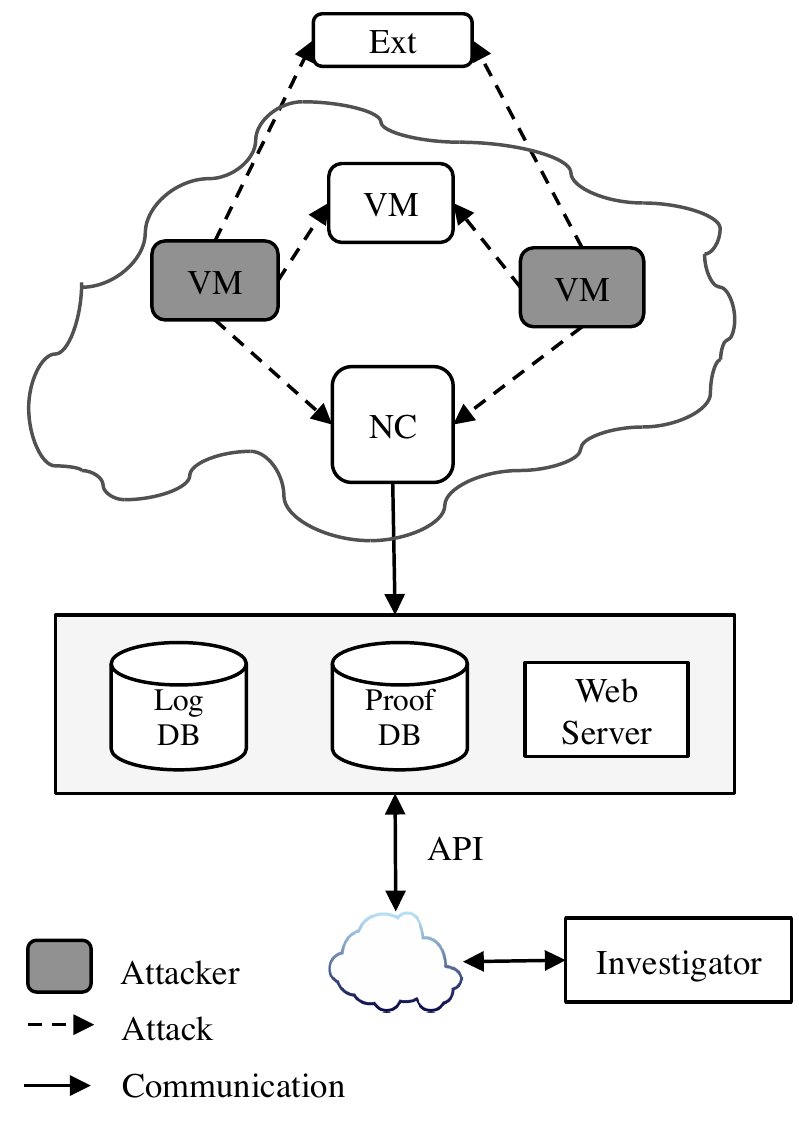}
\caption{Overview of SecLaaS}
\label{figure:overview}
\end{figure}

A VM in the cloud can attack other VMs inside the cloud or can attack a computing device outside the VM. The attacker VM can also attack the \emph{Node Controller} (NC) to launch a side channel attack \cite{ristenpart2009hey}. Figure \ref{figure:overview} presents an overview of storing the logs in a secured way and making it available to forensic investigators in case of such attacks.
Malicious activity of a VM can be found from various logs generated in the NC, on which the VM is running. For each running VM, our system first extracts various kinds of logs from the NC and will store in a persistent log database. Hence, terminating the VM will not prevent SecLaaS to provide useful logs during investigation. While saving logs in log database, SecLaaS ensures the integrity and confidentiality of the logs. After saving a log entry in the log database, the system will additionally store the proof of this entry in the proof database. When an investigator wants logs of a particular IP to investigate an incident, he can get the necessary logs by an API call or from the cloud management plane. In order to prove the logs as admissible evidence, the investigator can provide the proof of the logs along with the logs.

\subsection{Schematic Description}
In this section, we present the schematic description of our system. SecLaaS extracts log information from different log sources and generates a Log Entry \textit{LE}. A Log Entry \textit{LE} for network log is defined as follows:
\begin{equation}
	LE = <FromIP,ToIP,T\textsubscript{L},Port,UserId>
\end{equation}
To ensure the confidentiality of users' log, some information of the \textit{LE} can be encrypted using a common public key of the security agencies. The Encrypted Log Entry \textit{ELE} is prepared as follows:
\begin{equation}
	ELE = <E_{P_{K_a}}(ToIP,Port,UserId),FromIP, T\textsubscript{L}>
\end{equation}
where $P_{K_a}$ is the common public key of all the agencies. We cannot encrypt all the fields of the \textit{LE} as the CSP needs to search the storage by some fields.
To preserve the correct order of the log entry, we will use a hash-chain scheme. We refer the hash-chain as Log Chain (\textit{LC}), which will be generated as follows:
\begin{equation}
	LC = <H(ELE,LC\textsubscript{Prev})>
\end{equation}
where LC\textsubscript{Prev} is the Log Chain \textit{LC} of the previous entry of the persistent storage.
Each entry for the persistent log database \textit{DBLE} is constituted of ELE and LC,
\begin{equation}
	DBLE = <ELE,LC>
\end{equation} 
The proof of this \textit{DBLE} will be inserted into a accumulator. We denote this as Accumulator Entry \textit{AE}.
At the end of each day, CSP retrieves the \textit{AE\textsubscript{D}} of that day and generates the Proof of Past Log \textit{PPL} as follows:
\begin{equation}
	PPL = <H(AE\textsubscript{D}),S\textsubscript{PKc}(AE\textsubscript{D}),t>
\end{equation}
where \textit{H(AE)} is the hash of \textit{AE}, t represents the proof generation time, and \textit{S\textsubscript{PKc}(AE)} is the signature over \textit{AE} using the private key of the CSP, PKc.

\subsection{System Details}
In this section, we present how the log insertion, proof generation, and verification of SecLaaS work. We consider the network log to describe the entire system. After generating the Log Entry \textit{LE} the system will work for any type of logs.\smallskip

\subsubsection{Log and Proof Insertion} 
\begin{figure}[!ht]
\centering
\includegraphics[width=0.48\textwidth]{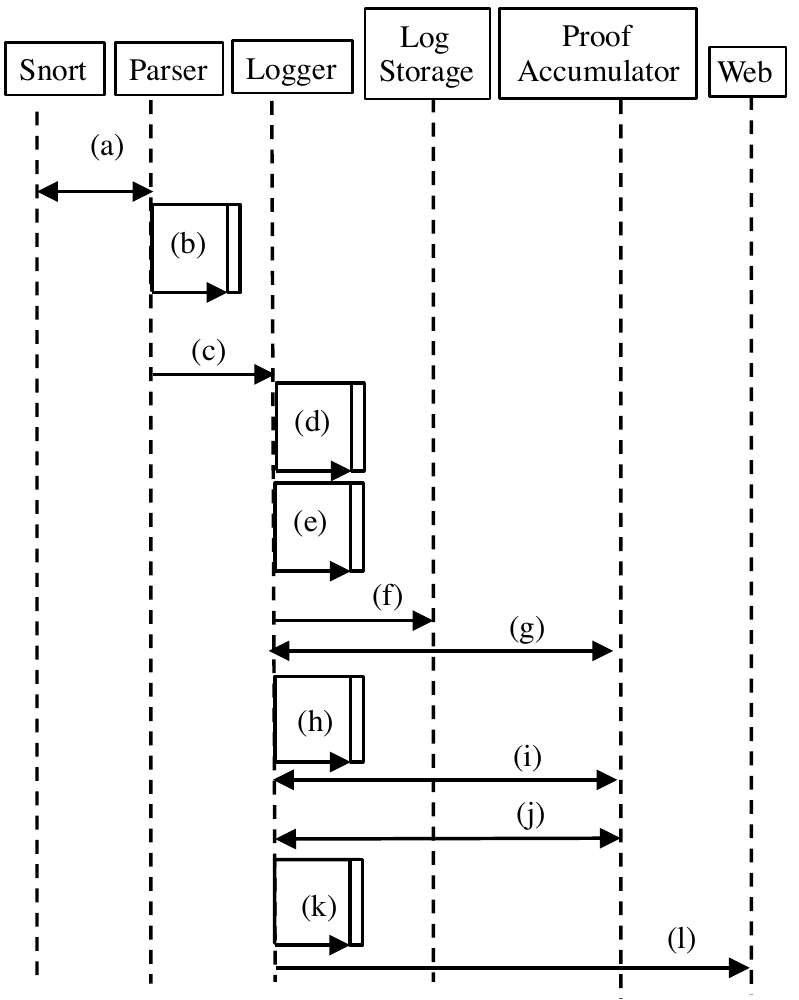}
\caption{Process Flow of Retrieving Log and Storing the PPL }
\label{figure:flow}
\end{figure}

Figure \ref{figure:flow} illustrates the detail flow of log retrieval, secured log insertion, and PPL generation and below is the details of the system.
\renewcommand{\labelenumi}{(\alph{enumi})}
\begin{enumerate}
	\item The parser module first communicates with the log sources to collects different types of logs. For example, to store network log the parser listens the Snort \footnotemark[1]. 
\footnotetext[1]{http://www.snort.org}	
	
	\item After acquiring logs from different sources, the parser then parses the collected log and generates the Log Entry \textit{LE}.
	
	\item The parser module sends the Log Entry \textit{LE} to the logger module to further process the \textit{LE}.
	
	\item The logger module, upon receiving the \textit{LE} from the parser, encrypts some confidential information using the public key of the security agencies and generates the Encrypted Log Entry \textit{ELE}. The private key to decrypt the log can be shared among the security agencies. For the network log, some crucial information that we can encrypt includes: destination IP, port, and user information. 
	
	\item After generating the \textit{ELE}, the logger module then creates the Log Chain \textit{LC} by using equation 3. In section \ref{sec:security}, we will discuss how this can prevent reordering and deletion of logs.
	
	\item The logger module then prepares the entry for the log storage \textit{DBLE} using the Encrypted Log Entry \textit{ELE} and the Log Chain \textit{LC}, and sends the \textit{DBLE} to the log storage to add the new entry.

	\item After creating the database entry \textit{DBLE}, the logger module communicates with the proof storage to retrieve the latest accumulator entry.
	
	\item In this step, the logger generates the proof of the database entry \textit{DBLE}, i.e., the logger creates a new entry for the accumulator \textit{AE} and updates the last retrieved accumulator entry with the newly generated \textit{AE}.
	
	\item The logger module sends the updated accumulator entry to the accumulator storage to store the proof.
	
	\item At the end of each day, the logger retrieves the last accumulator entry of each static IP, which we denote as AE\textsubscript{D}.
	
	\item According to equation 5, the logger then creates the Proof of Past Log \textit{PPL} using the AE\textsubscript{D}.
	
	\item After computing the Proof of Past Log \textit{PPL}, the logger will publish the \textit{PPL} and the public key of CSP on the web. These information can also be available by RSS feed to protect it from manipulation by the CSP after publishing the proof. We can also build a trust model by engaging other CSPs in the proof publication process. Whenever one CSP publishes a PPL, that PPL will also be shared between other CSPs. Therefore, we can get a valid proof as long as one CSP is honest.
	
\end{enumerate}

\subsubsection{Verification} 
When an investigator wants to investigate an incident, he will first gather the required log either by calling log API or from the cloud management console. While presenting the evidence to the court, he needs to provide the collected logs and also the proof of the logs. There will be two steps to verify the provided logs. In the first step, the auditor will verify the integrity of the proof and the individual log entry. In the next step, he will verify the order of the log.  \smallskip

\begin{figure}[!ht]
\centering
\includegraphics[width=0.48\textwidth]{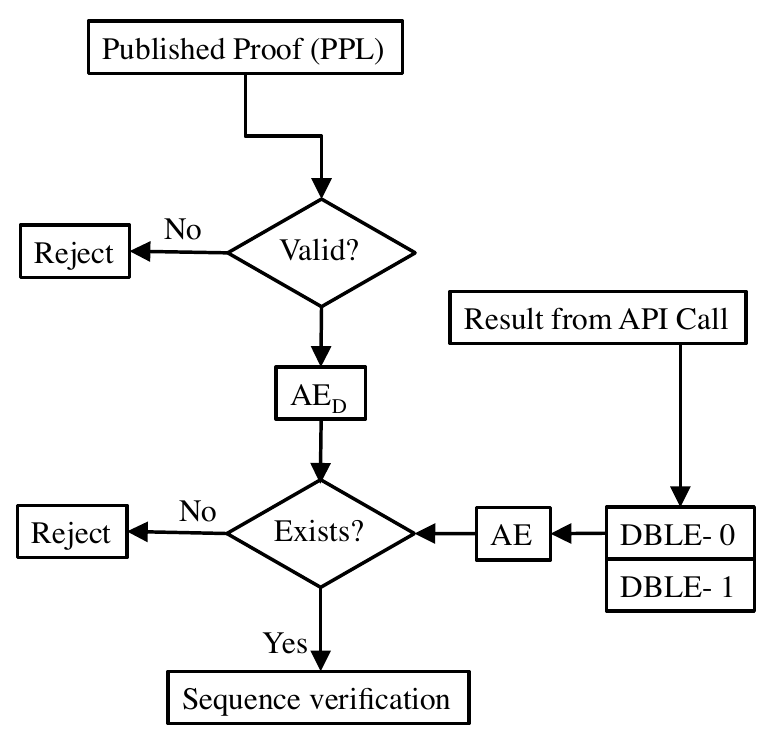}
\caption{Log Verification Process Flow}
\label{figure:validity}
\end{figure}

\noindent\textbf{Integrity Verification: } Figure \ref{figure:validity} shows the the process flow of individual log entry verification.

The verification process starts from checking the validity of the published Proof of Past Log \textit{PPL}. To do so, first, the auditor decrypts the S\textsubscript{PKc}(AE\textsubscript{D}) using the public key of the CSP and he will get the AE\textsubscript{D}. Then the auditor generates the hash value from the dycrypted AE\textsubscript{D}. If the generated hash and the H(AE\textsubscript{D}) of the \textit{PPL} matches, then the auditor accepts the \textit{PPL} as a valid proof of log, otherwise he rejects the verification process.

In the next step, the auditor generates the Accumulator Entry \textit{AE} for each \textit{DBLE}. Then, he will check whether the calculated \textit{AE} exists in the AE\textsubscript{D}. If exists, then the auditor proceeds towards log order verification process, otherwise he rejects the provided log information. \smallskip

\begin{figure}[!ht]
\centering
\includegraphics[width=0.48\textwidth]{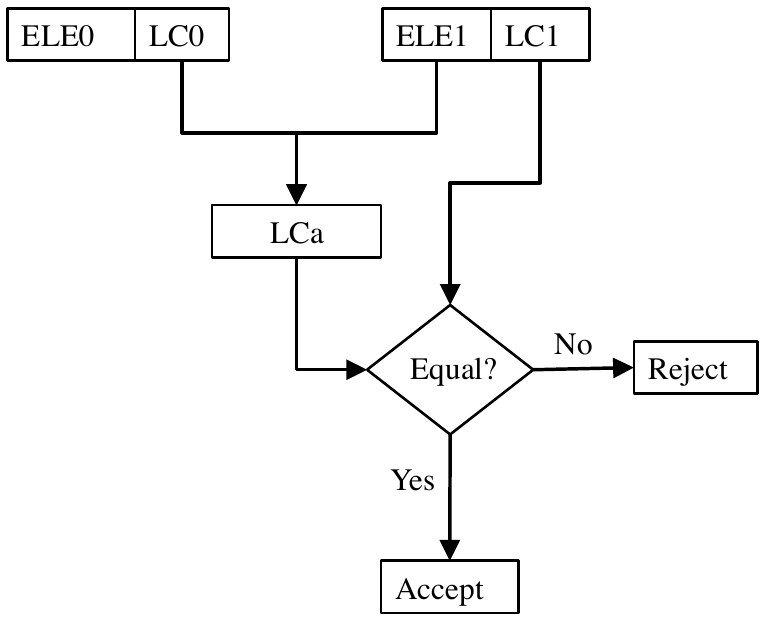}
\caption{Log Order Verification Process Flow}
\label{figure:order}
\end{figure}

\noindent\textbf{Sequence Verification: } Figure \ref{figure:order} illustrates the log order verification process, where we verify whether the current log (\textit{DBLE1}) is actually after the previous log (\textit{DBLE0}) in the original sequence of log generation. In the figure \ref{figure:order}, \textit{ELE0} denotes the Encrypted Log Entry \textit{ELE} of the first log and \textit{ELE1} represents the same for the second log. To verify the correct order, the auditor  calculates the Log Chain \textit{LCa} from the first Log Chain \textit{LC0} and the second Encrypted Log \textit{ELE1} according to the following equation.
\begin{equation}
	LCa = <H(ELE1,LC0)>
\end{equation}
If \textit{LCa} matches with the 2nd Log Chain \textit{LC1} then the auditor accepts the logs, otherwise he rejects it.

\section{Security Analysis}
\label{sec:security}
As CSPs have control over generating the logs and the proofs, they can always tamper with the logs. After acquiring logs through API or management console, investigators can also alter the logs before presenting it to court. Therefore, here we propose a tamper evident scheme. Any violation of the integrity and confidentiality properties, as mentioned in Section \ref{sec:threatmodel} can be detected during the verification process.

In our collusion model, there are three entities involved -- CSP, user, and investigator. All of them can be malicious individually or can collude with each other. We denote an honest CSP as C, a dishonest CSP as $\bar{C}$, an honest user as U, a dishonest user as $\bar{U}$, an honest investigator as I, and a dishonest investigator as $\bar{I}$. Hence, there can be total eight possible combinations of collusion. Table \ref{table:collusion} presents all the combinations of collusion, possible attacks for each collusion, and required security properties to defend that collusion.  Here, we discuss how our proposed system can ensure all the security properties, which are required to protect collusion between CSP, user, and investigator. 
\begin{table*}
\begin{center}
	\renewcommand{\arraystretch}{1.25}
	\begin{tabular}{|p{0.03\textwidth} | p{0.03\textwidth} | p{0.09\textwidth} | p{0.06\textwidth} | p{0.45\textwidth}| p{0.16\textwidth}|}
	\hline
	\multicolumn{3}{|c|}{\textbf{Is Honest?}} & \textbf{Notation} & \textbf{Attack} & \textbf{Required Security \newline Properties}\\ \hline	
	\textbf{CSP} & \textbf{User} & \textbf{Investigator} & &  & \\ \hline	
	\ding{51} & \ding{51} & \ding{51} & C U I & No attack & None\\ \hline
	\ding{53} & \ding{51} & \ding{51} & $\bar{C}$ U I & Reveal user activity from logs & C2 \\ \hline
	\ding{51} & \ding{53} & \ding{51} & C $\bar{U}$ I & Recover other cloud users' log from published proof & C1\\ \hline
	\ding{51} & \ding{51} & \ding{53} & C U $\bar{I}$ & Remove, reorder, and plant fake logs & I4, I5, I6 \\ \hline
	\ding{51} & \ding{53} & \ding{53} & C $\bar{U} \bar{I}$ & Remove, reorder, and plant fake logs & I4, I5, I6 \\ \hline
	\ding{53} & \ding{51} & \ding{53} & $\bar{C}$ U $\bar{I}$ & Remove, reorder, plant fake logs, and repudiate published PPL & I1, I2, I3, I4, I5, I6, I7 \\ \hline
	\ding{53} & \ding{53} & \ding{51} & $\bar{C} \bar{U}$ I & Remove, reorder, plant fake logs, and repudiate published PPL & I1, I2, I3, I7\\ \hline
	\ding{53} & \ding{53} & \ding{53} & $\bar{C} \bar{U} \bar{I}$ & Remove, reorder, plant fake logs, and repudiate published PPL & I1, I2, I3, I4, I5, I6, I7\\ \hline
	\end{tabular}

\end{center}
\caption{Collusion model, possible attacks and required security properties}
\label{table:collusion}
\end{table*}
\begin{itemize}
	\item \textit{I1, I2, I4, I5}: A CSP can collude with the cloud user or the investigator and can remove crucial log information. Also, while providing logs through the API or the management console, the CSP can simply hide some crucial log entries. An Investigator can also hide logs before at the time of presenting evidence to court, though he have received correct logs through the log API. However, at the verification stage, our system can detect any such removal of log entries. Let us assume that there are three log entries DBLE0, DBLE1, and DBLE2 and their proof has already been published. Now, if CSP removes DBLE1 and provides only DBLE0 and DBLE2 to the investigator, then this removal can be easily detected at the sequence verification stage. In this case, the hash of LC0 and ELE2 will not match with the LC2 because the original LC2 was calculated by hashing LC1 and ELE2. 
	In the same way, an auditor can detect the re-ordering of logs. For example, while providing the logs to an auditor, if the CSP or investigator provides the log in DBLE0, DBLE2, DBLE1 order, then using the same technique, the auditor can identify that DBLE2 does not come after DBLE0 in actual generation order. A CSP can further try to change the DBLE2 by replacing the original LC2 with a new Log Chain value so that, in the sequence verification process, the order breaking will not be detected. However, an attempt of changing the DBLE2 will be detected during the individual log entry verification phase. The accumulator entry of the fake DBLE2 will not exist in the published Proof of Past Log PPL.

	\item \textit{I3, I6}: A colluding CSP can plant false log information while providing the log to the investigator. However, if the CSP does this after publishing the proof, our system can detect these phony logs. A dishonest investigator can also try to frame an honest user by presenting fake logs to the court. Suppose, DBLE\textsubscript{F} is the fake log and the auditor generates the Accumulator Entry AE\textsubscript{F} for this log. If it's fake, then AE\textsubscript{F} will not be present in the AE\textsubscript{D} of the Proof of Past Log \textit{PPL} and the auditor can reject that incorrect log.
	
	\item \textit{I7}: After publishing the proof of past log \textit{PPL}, CSP cannot repudiate  the published proof as the accumulator entry \textit{AE\textsubscript{D}} is signed by CSP's private key. Nobody other than the CSP can use that private key to sign the \textit{AE\textsubscript{D}}. Hence, after decrypting the signed value and generating hash on the decrypted value, if it matches with the hashed \textit{AE\textsubscript{D}} value, the CSP cannot repudiate the published value. Additionally, if the CSP comes up with a false \textit{PPL\textsubscript{f}} in place of a published \textit{PPL}, then it will be easily detected. In that case, the \textit{H(AE\textsubscript{D})} of the published \textit{PPL} and the \textit{H(AE\textsubscript{Df})} of the false \textit{PPL\textsubscript{f}} will not be same. As the CSP has already signed the \textit{AE\textsubscript{D}} of the published \textit{PPL} using its private key, it cannot deny the published value.
	
	\item \textit{C1, C2}: To store the proof of the logs, we propose to use an accumulator function, which will ensure the C1 property, i.e., from the proof of logs, adversaries cannot recover any log. We implement our scheme using Bloom filter and One-Way Accumulator, which can ensure this property.
	While storing the log data in persistent storage, we propose to encrypt some crucial information e.g., user id, destination IP, etc by using a common public key of all the investigator agencies. Hence, a malicious cloud employee cannot retrieve plain log information from the persistent storage; e.g., identifying the visiting IPs of a particular user will not be possible by the malicious cloud employee. In this way, our scheme can ensure the C2 property.
\end{itemize}
\section{Implementation and Evaluation}
\label{sec:implementation}
In this section, we present the implementation of SecLaaS on OpenStack and performance analysis of the scheme using different types of accumulators.

\subsection{Implementation}
\begin{figure}[!ht]
\centering
\includegraphics[width=0.48\textwidth]{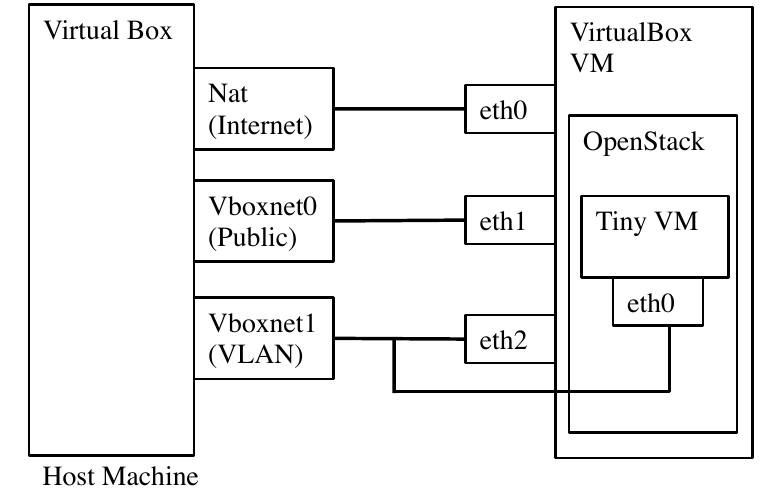}
\caption{Prototype Environment Configuration}
\protect\cite{websitetikalopenstack}
\label{figure:openstack}
\end{figure}

\noindent\textbf{System Setup:} 
We used Openstack\footnotemark[1] and Snort for testing and implementation of our project. OpenStack is an open source cloud computing software and Snort is a free lightweight network intrusion detection system. We created the virtual environments with VirtualBox (a free virtualization software)\footnotemark[2] running on a single Ubuntu machine. Figure \ref{figure:openstack} illustrates the system setup and below is the description of the system:

\footnotetext[1]{http://www.openstack.org}
\footnotetext[2]{https://www.virtualbox.org}

\begin{itemize}

\item Host machine's hardware configuration: Intel Core I7 quad core CPU, 16 GB ram and 750 GB hard drive. Ubuntu 12.04 LTS 64-bit is used as Host Operating System.

\item VirtualBox  4.1.22 r80657 for Ubuntu 12.04 LTS

\item Openstack (Essex release, came out by the end of April 2012) installation in VirtualBox; for simplicity, the system had one node controller. Configuration of vitualized cloud controller: Intel 2.5Ghz Dual Core cpu, 8 GB ram and 20 GB hard drive. Ubuntu 12.04 64-bit Sever edition is used as the Operating system for Openstack setup.

\item In the virtualized environment, the Cloud Controller required following network adapter configuration in VirtualBox to work properly:

\begin{itemize}
	
	\item Adapter 1: Attached to NAT- eth0 of the Cloud controller is connected here.
	
	\item Adapter 2: Host-only network for Public interface- connected with eth1 (IP was set to 172.16.0.254, mask 255. 255.0.0, dhcp disbaled)
	
	\item Adapter 3: Host-only network for Private (VLAN) interface connected with eth2 (IP to 11.0.0.1, mask 255. 0.0.0, dhcp disbaled)

\end{itemize}
\item We used RSA (2048 bit) for signature generation and SHA-2(SHA-256) hash function for hashing.
\end{itemize}

We set up Snort in node controller to track the network activity of the virtual machines. We added two virtual machines: the first one had private IP: 11.1.0.3 and public IP: 172.16.1.1; while the other had private IP: 11.1.0.5 and public IP: 172.16.1.3. Here is a sample Snort log:

\begin{center}
``11/19-13:43:43.222391 11.1.0.5:51215 -> 74.125.130.106:80 \ 
TCP TTL:64 TOS:0x0 ID:22101 IpLen:20 DgmLen:40 DF \
***A***F Seq: 0x3EA405D9  Ack: 0x89DE7D  Win: 0x7210  TcpLen: 20''
\end{center}

\begin{figure}[!ht]
\centering
\includegraphics[width=0.48\textwidth]{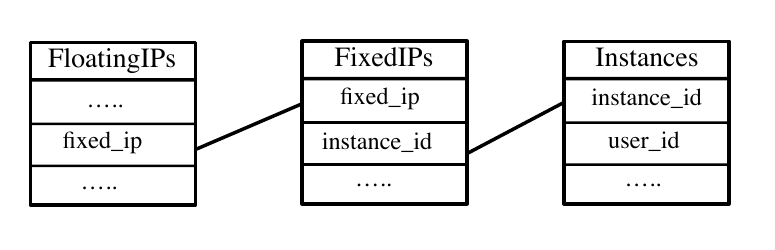}
\caption{Resolving User ID from Public IP}
\label{figure:dbrelation}
\end{figure}

This log tells that the virtual machine with private IP 11.1.0.5 performed a http request to machine 74.125.130.160. By reverse engineering Openstack's ``nova'' mysql database, it is also possible to find out the static private IP and user information from a public IP. We used the references among FloatingIps, FixedIps and Instances tables to resolve the user id for a particular log entry. Figure \ref{figure:dbrelation} shows the relation between these three tables. 

We implemented the Proof of Past Log \textit{PPL} scheme of the SecLaaS using two accumulators. One is BloomFilter \cite{bloom1970space} and another is One-Way Accumulator \cite{DBLP:conf/eurocrypt/93}. The steps from (g) to (k) will work differently for the two accumulators.\smallskip

\noindent\textbf{BloomFilter:} 		 	 	 						
A Bloom filter is a probabilistic data structure with no false negatives rate, which is used to check whether an element is a member of a set or not \cite{bloom1970space}. Bloom filter stores the membership information in a bit array. Bloom filters decrease the element insertion time and membership checking time. The only drawback of the Bloom filter is the probability of finding false positives. However, we can decrease the false positive probability by using a large bit array. 

To use the Bloom filter as a proof, we use one bloom filter for one static IP for each day. That means, one Bloom filter stores the proof of all the logs of one static IP for a particular day. In step (g), the logger retrieves the bloom filter from the proof storage, which holds the bit positions for the previously inserted logs of the day. In step (h), while creating the accumulator entry \textit{AE}, the logger will generate the k number of bit positions for the database entry \textit{DBLE} by hashing the log for k times. Then, the logger updates the previously retrieved Bloom filter with the newly generated \textit{AE} and sends the updated Bloom filter to the proof storage. At the end of each day, the CSP will retrieve the Bloom filter entry of each static IP \textit{AE\textsubscript{D}} and create the proof of past log PPL for that day using equation 5.

In the verification phase, after verifying the validity of the published proof, the auditor will hash the log entry that he has received from the API call and calculate the bit positions of the Bloom filter. Then he will compare these bit positions with the published \textit{AE\textsubscript{D}}. If all the calculated bit positions are set in the published Bloom filter \textit{AE\textsubscript{D}}, then the verifier will be sure about the validity of the log. One single false bit position means the log entry is not valid. \smallskip

\noindent\textbf{One-Way Accumulator:}
A One-Way accumulator is a cryptographic accumulator, which is based on RSA assumption and provides the functionality of checking the membership of an element in a set \cite{benaloh1994one}. This scheme works with zero false negative and false positive probability. Initially, we create the public and private values for the accumulator. The private values are two large prime numbers P and Q. The first public value is N, where N = P*Q and the second public value is a large random number which is the initial seed X. 

In step (g), the logger retrieves the accumulator entry \textit{AE}. If there is no proof entry, i.e., the \textit{AE} is empty for an IP on a day, then the \textit{AE} of the first \textit{DBLE} of the day is generated using the following equation
\begin{equation}
	AE = X\textsuperscript{H(DBLE)} mod N
\end{equation}
where \textit{H(DBLE)} is a numeric hash value of \textit{DBLE}. If the retrieved \textit{AE} is not empty, then the new \textit{AE} will be generated using the following equation
\begin{equation}
	AE = AE\textsuperscript{H(DBLE)} mod N
\end{equation}
The logger module then sends the calculate \textit{AE} to the proof storage.

At the end of the day, the logger retrieves the last accumulator entry \textit{AE\textsubscript{D}} and creates the proof of past log \textit{PPL} for the day using equation 5. The logger needs to do some additional computation here comparing with the Bloom filter. It will generate an identity for each \textit{DBLE} and tagged it with the \textit{DBLE}. If there are k number of \textit{DBLE} on a day then the identity \textit{ID} of the i\textsuperscript{th} DBLE will be calculated using the following equation
\begin{equation}
	ID = X\textsuperscript{H(DBLE\textsubscript{1}) H(DBLE\textsubscript{2})...
	H(DBLE\textsubscript{i-1}) H(DBLE\textsubscript{i+1})....H(DBLE\textsubscript{k})}
\end{equation}

While verifying the validity of the DBLE\textsubscript{i}, the verifier computes ID\textsuperscript{H(DBLE\textsubscript{i})}mod \textit{N} and compares it with AE\textsubscript{D}.
If AE\textsubscript{D} = ID\textsuperscript{H(DBLE\textsubscript{i})}mod \textit{N}, then the verifier will be sure about the validity of the log.
  
\subsection{Evaluation}
 
To evaluate the performance of our scheme we ran our experiment using multiple accumulators. For Bloom filter, we used two configurations: 1\% false positive (FP) probability with 5000 items, and 2\% FP with 10000 items. Then, for One-Way accumulator, we choose 32-bit and 64-bit P,Q, and X. 

\begin{figure}[!ht]
\centering
\includegraphics[width=0.48\textwidth]{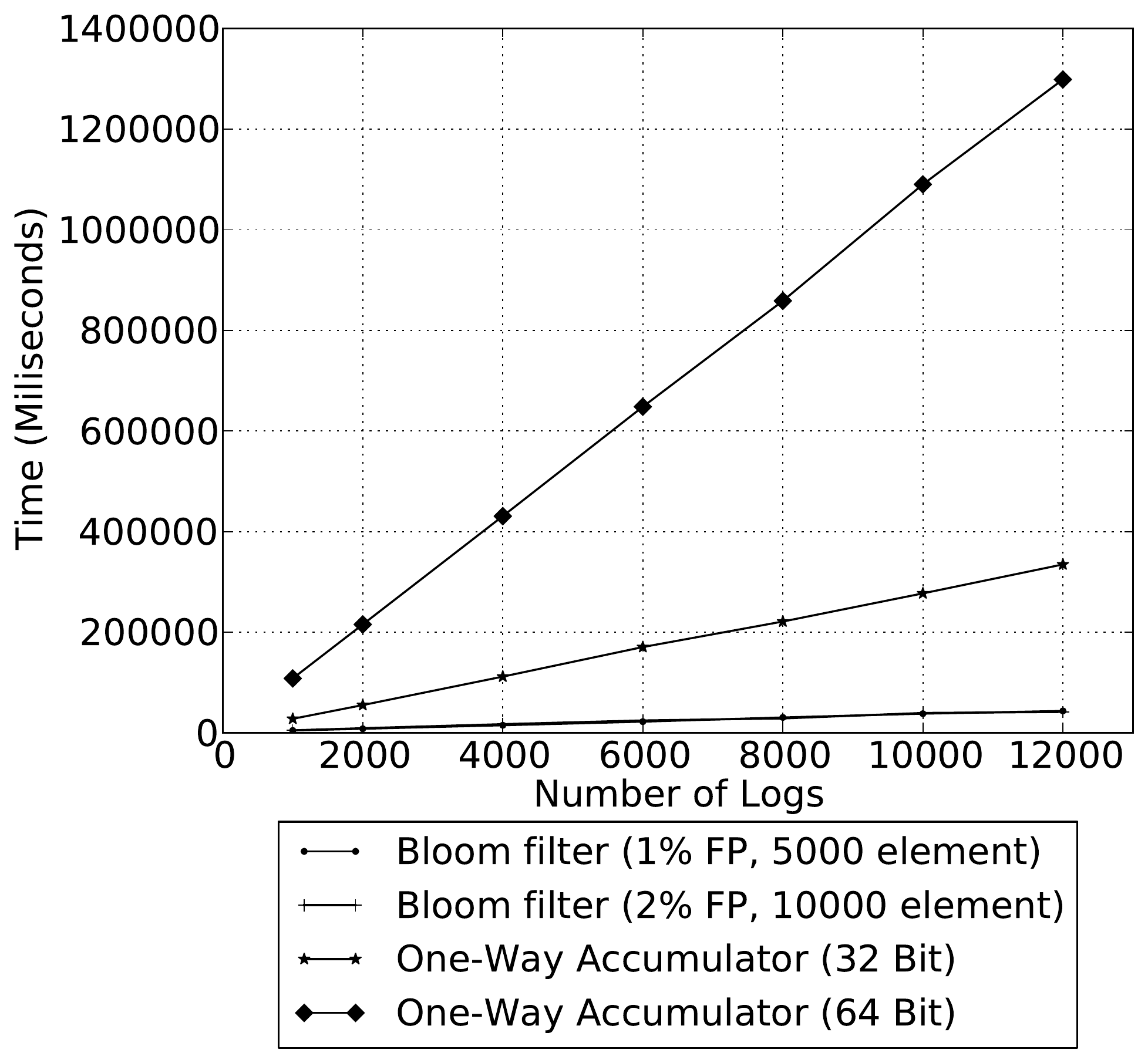}
\caption{Performance Analysis of Log Insertion Using Different Accumulators }
\label{figure:insertion}
\end{figure}

Figure \ref{figure:insertion} shows the performance analysis of log insertion including generating and updating the proof of log, i.e., the time required to complete the steps from  (b) to (i) of the Figure \ref{figure:flow}. We found that for all the accumulators, time increases linearly with the increase of log size. For the two Bloom filter configurations, we noticed nearly similar time. However, we found a significant amount of increase in time when changed the one-way accumulator from 32-bit to 64-bit.
  
\begin{figure}[!ht]
\centering
\includegraphics[width=0.48\textwidth]{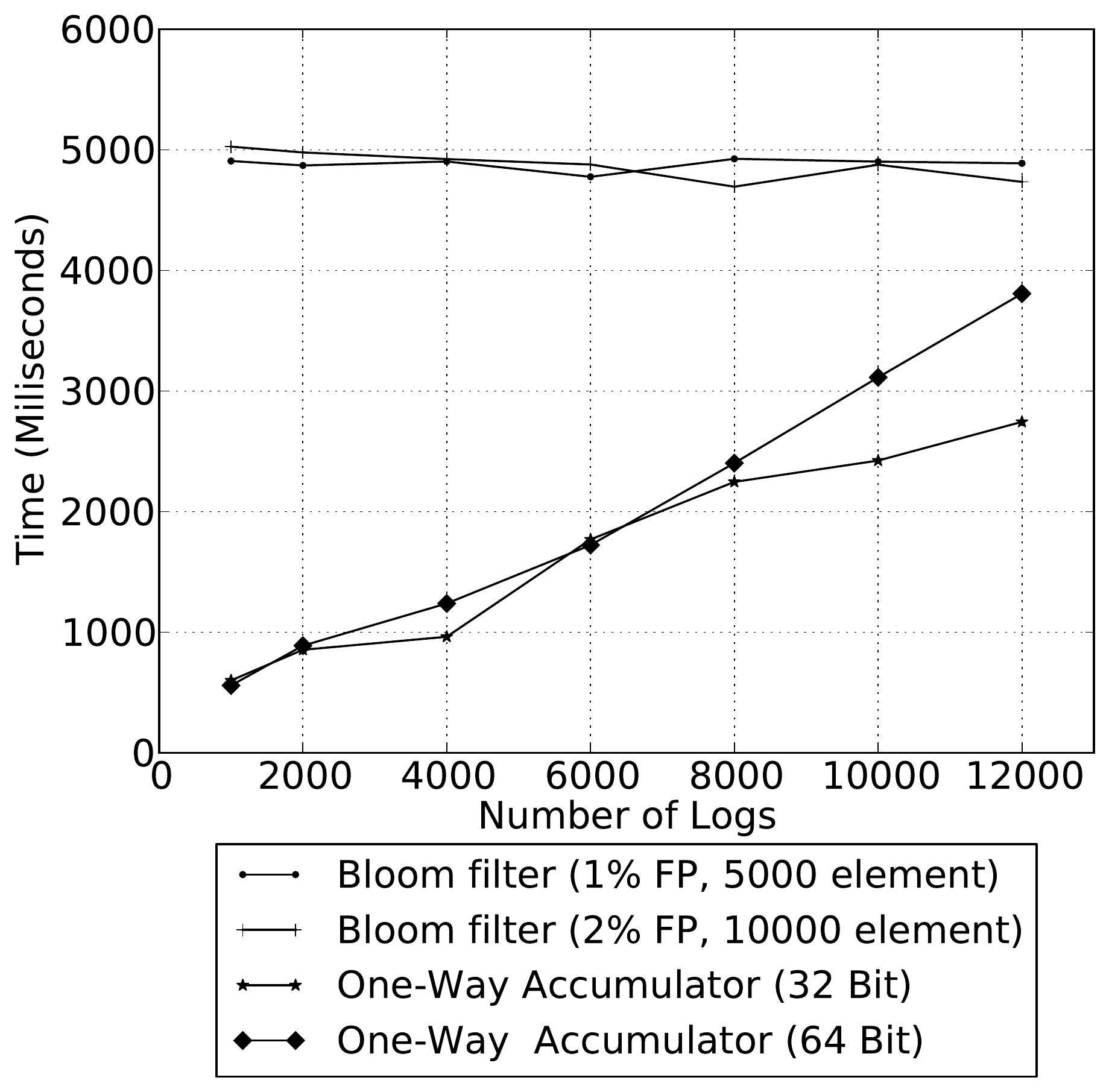}
\caption{Performance Analysis of PPL Generation Using Different Accumulators }
\label{figure:ppl}
\end{figure}

Figure \ref{figure:ppl} illustrates the performance analysis of generating the Proof of Past Log of a day for different accumulators, i.e., the time required to complete the steps from (j) to (l) of the figure \ref{figure:flow}. For the two different Bloom filters, we found nearly constant amount of time. On the contrary, for the One-Way accumulators, we found a linear increase in time for different sizes of logs. It is obvious, because in the later case, we need to compute the identity of each log entry using the equation 9 which has \textit{O(n)} time complexity. \cite{goodrich2002efficient}.

\begin{figure}[!ht]
\centering
\includegraphics[width=0.48\textwidth]{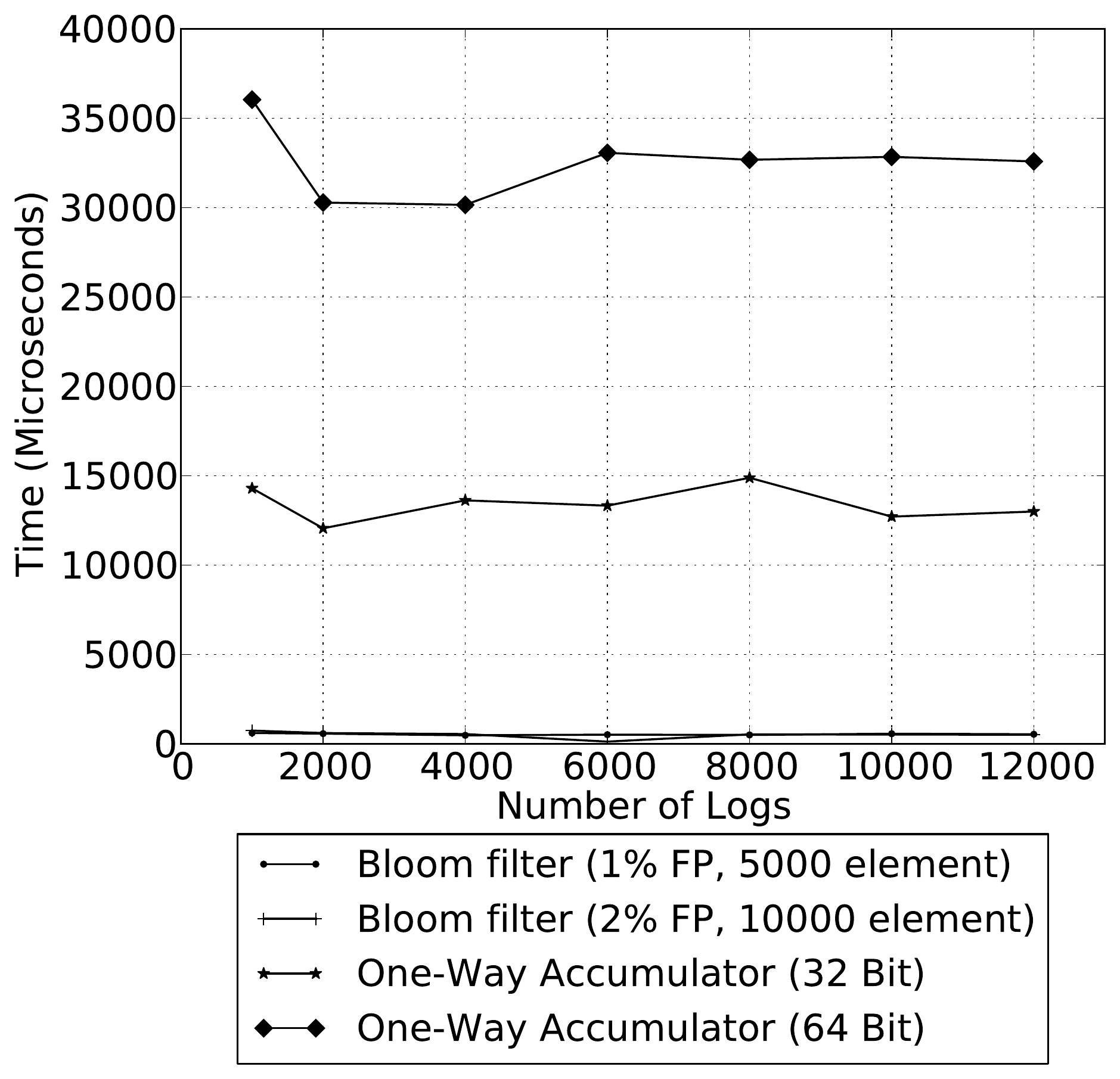}
\caption{Performance Analysis of Log Verification Using Different Accumulators }
\label{figure:verify}
\end{figure}

Figure \ref{figure:verify} presents the time for verifying the validity of each log. For all of the accumulators, we found nearly constant amount of time with the increase in log size. However, the time required for 32-bit accumulator is higher than the Bloom filters and for 64-bit accumulator the time is significantly higher than its counterparts.

To identify the performance degradation of NC for storing log, we first run a RSA encryption on a 16 MB file for several times and measure the average execution time without running the snort logger in NC. At that time, two VMs were running on that NC. Then we start the snort service and again measure the average execution time of encryption on the same data. From these two execution times we measured the performance overhead, which is only 1.6\%.

\section{Discussion}
\label{sec:discussion}
Our experimental result shows that the Bloom filter outperforms the One-Way accumulator for all the tasks: log insertion, \textit{PPL} generation, and log verification. However, Bloom filter is a probabilistic accumulator, which can state about the existence of a log in the \textit{PPL} with certain probability. It works with zero false negative probability though. On the other hand, One-Way accumulator works with zero false positive probability. This means, in Bloom filter there is still some chance of planting false log information by the CSP or the investigator, which is not possible in One-Way accumulator. The later one always finds a valid log entry with zero false positive probability. However, we can decrease the false positive probability of the Bloom filter by allocating more space to the bit array. For example, to ensure 1\% FP for 10,000 elements we need 91133 bits or 11.12 KBytes storage and to ensure 0.1\% FP for the same number of elements we need 111945 bits or 13.67 KBytes storage. In our scheme we use one Bloom filter for one static IP for each day. If we have n number of static IP then for 0.1\%  FP and 10,000 logs we will require n * 13.67 KBytes storage each day and n*4989.55 KBytes in one year. 

For the One-Way accumulator, proof requires a very small amount of storage. The 32-bit accumulator requires 19 Bytes for the final accumulator entry, whereas, the 64-bit requires 39 Bytes. However, to complete the verification in O(1) time, we need to pre-compute the identity of each log record and store it along with the logs. For a 32-bit accumulator, we require 10 Bytes for each identity and for the 64-bit the requirement is 20 Bytes. That means, for 10,000 records, we need 97.67 KBytes storage with the 32-bit accumulator and 195.35 KBytes for the 64-bit accumulator. Hence, we get a 600\% increase in storage in one year for the 32-bit accumulator comparing with the 0.1\% FP for 10,000 records. However, this extra storage can provide us zero false positive probability. Therefore, we need to choose whether we will go for accommodating higher storage with the One-Way accumulator or tolerating a little false positive probability with the Bloom filter. Moreover, the Bloom filter will give us better performance in all the required tasks.
\section{Related Works}
\label{sec:relatedwork}
As logging information is one of the prime needs in forensic investigation, several researchers have explored this problem across multiple dimensions. Marty proposed a log management solution, which can solve several challenges of logging, discussed in Section~\ref{sec:background} \cite{marty2011cloud}. In his solution, after enabling logging on all infrastructure components to collect logs, he proposed to establish a synchronized, reliable, bandwidth efficient, and encrypted transport layer to transfer log from the source to a central log collector. Final step deals with ensuring the presence of the desired information in the logs. The proposed guideline tells us to focus on when to log, what to log, and how to log. The answer of when to log depends on the use-cases, such that business relevant logging, operations based logging, security (forensics) related logging, and regulatory and standards mandates. At minimum, he suggested to log the time-stamps record, application, user, session ID, severity, reason, and categorization, so that we can get the answer of what, when, who, and why (4 W). However, this work does not provide any solution for logging network usage, file metadata, process usage, and many other important sources of evidence. 
	
As a solution of forensic investigation, Zafarullah et al. proposed logging provided by OS and the security logs \cite{zafarullah2011digital}. In order to investigate the digital forensics in cloud, they set up cloud computing environment using Eucalyptus. Using Snort, Syslog, Log Analyzer (e.g., Sawmill), they were able to monitor the Eucalyptus behaviour and log all internal and external interaction of Eucalyptus components. For their experiment, they launched a DDoS attack from two virtual machine and analyzed bandwidth usage log and processor usage log to detect the DDoS attack. From the logs in \emph{/var/eucalyptus/jetty-request-05-09-xx} file on Cloud Controller (CC) machine, it is possible to identify the attacking machine IP, browser type, and content requested. From these logs, it is also possible to determine the total number of VMs controlled by a single Eucalyptus user and the VMs communication patterns. Their experiment shows that if the CSPs come forward to provide better logging mechanism, cloud forensics will get a better advancement.  

To make the network, process and access logs available to the customer, Bark et al. proposed exposing read-only API by the CSP \cite{birk2011technicalIssues}. By using these APIs, customer can provide valuable information to investigator. In PaaS, customers have full control on their application and can log variety of access information in a configurable way. So for PaaS, they proposed a central log server, where customer can store the log information. In order to protect log data from possible eavesdropping and altering action, customers can encrypt and sign the log data before sending it to the central server. In the same context, Dykstra et al. recommended a cloud management plane, for using in IaaS model \cite{dykstraacquiring}. From the console panel, customers, as well as investigators can collect VM image, network, process, database logs, and other digital evidence, which cannot be collected in other ways. 

Secure logging has been discussed in several research works \cite{Ma:2009:NAS:1502777.1502779,rafaelsecremotelog2006,Schneier:1999:SAL:317087.317089}. However, none of these works focus on secure logging in cloud environment, specially providing secure logging as a service. Moreover, they did not consider the logger as dishonest. In the threat model of current secure logging works, researchers consider  attacks on privacy and integrity from external entity. These works do not consider collusion between different entities. The closest work that we can relate to our work is a secure logging scheme proposed by Yavuz et al., which provides public verifiability of audit logs for distributed system \cite{Yavuz2009baf}. Using their proposed scheme, time required for logging and verification increase with the number of logs. On the other hand, in our system, time required for log verification is almost constant with number of logs using various types of accumulators.

The solution proposed by Marty provided a guideline for logging criteria and answered some importation questions, e.g., what are the information we need to log, how to log and when to log. Zafarullah et al. showed that it is possible to collect necessary logs from cloud infrastructure, while Bark et al. and Dykstra et al. proposed for public API or management console to mitigate the challenges of log acquisition. However, none of them proposed any scheme of storing the logs in Cloud and making it available publicly in a secure way. Dyskstra et al. mentioned that the management console requires an extra level of trust and the same should hold for APIs. In this paper, we took the first step towards providing a solution to mitigate these challenges. Combining all the previous solutions and our scheme will drive towards making the Cloud more forensics friendly. 
\section{Conclusion and Future Work}
\label{sec:conclusion}
Logs from different sources, e.g., network, process, database are a crucial source of evidence for forensics investigation. However, collecting logs from cloud is challenging as we have very little control over clouds compared to traditional computing systems. Till now, investigators need to depend on the CSP to collect logs of different sources. To make the situation even worse, there is no way to verify whether the CSP is providing correct logs to the investigators or the investigators presenting valid logs to the court. Moreover, while providing the logs to the investigators, the CSPs need to preserve the privacy of the cloud users. Unfortunately, there has been no solution which can make the logs available to the investigators and at the same time, can preserve the confidentiality and integrity of the logs. In this paper, we proposed \textit{SecLaaS}, which can be the solution to store and provide logs for forensics purpose securely. This scheme will allow the CSP to store the logs while preserving the confidentiality of the cloud users. Additionally, an auditor can check the integrity of the logs using the Proof of Past Log \textit{PPL} and the Log Chain \textit{LC}. We ran our proposed solution on OpenStack and found it practically feasible to integrate with the cloud infrastructure.

	Preserving the logs and the proofs of the logs will also increase the auditability of cloud environment. Using our scheme, it is possible to store and provide any types of logs from which we can get all the activities of cloud users. Auditability  is a vital issue to make the cloud compliant with the regulatory acts, e.g., Sarbanes-Oxley (SOX) \cite{websitesarbanesox} or The Health Insurance Portability and Accountability Act (HIPAA) \cite{websitehippa}. Hence, implementing SecLaaS will make the cloud more compliant with such regulations, leading to widespread adoption of clouds by major businesses and healthcare organizations. 

	In future, we will integrate other logs besides the snort log with our proof-of-concept application. Moreover, we will continue experiment on different accumulators to find the best fitted accumulator algorithm with SecLaaS. And finally, we will implement SecLaaS as a module of OpenStack.

\section*{Acknowledgment}
This research was supported by a Google Faculty Research Award, the
Office of Naval Research Grant \#N000141210217, the Department of
Homeland Security Grant \#FA8750-12-2- 0254, and by the National
Science Foundation under Grant \#0937060 to the Computing Research
Association for the CIFellows Project.

\bibliographystyle{abbrv} 
\bibliography{cloudForensic}

\begin{thebibliography}{10}

\bibitem{rafaelsecremotelog2006}
R.~Accorsi.
\newblock On the relationship of privacy and secure remote logging in dynamic
  systems.
\newblock In {\em Security and Privacy in Dynamic Environments}, volume 201,
  pages 329--339. Springer US, 2006.

\bibitem{websiteamazon2009}
Amazon.
\newblock Zeus botnet controller.
\newblock
  \url{http://aws.amazon.com/security/security-bulletins/zeus-botnet-controller/}.
\newblock [Accessed July 5th, 2012].

\bibitem{websiteamazonwebservice}
AWS.
\newblock Amazon web services.
\newblock \url{http://aws.amazon.com}.
\newblock [Accessed July 5th, 2012].

\bibitem{benaloh1994one}
J.~Benaloh and M.~De~Mare.
\newblock One-way accumulators: A decentralized alternative to digital
  signatures.
\newblock In {\em Advances in Cryptology—EUROCRYPT’93}, pages 274--285.
  Springer, 1994.

\bibitem{birk2011technicalIssues}
D.~Birk and C.~Wegener.
\newblock Technical issues of forensic investigatinos in cloud computing
  environments.
\newblock {\em Systematic Approaches to Digital Forensic Engineering}, 2011.

\bibitem{bloom1970space}
B.~Bloom.
\newblock Space/time trade-offs in hash coding with allowable errors.
\newblock {\em Communications of the ACM}, 13(7):422--426, 1970.

\bibitem{websitehippa}
{Centers for Medicare and Medicaid Services}.
\newblock The health insurance portability and accountability act of 1996
  (hipaa).
\newblock \url{http://www.cms.hhs.gov/hipaa/}, 1996.
\newblock [Accessed July 5th, 2012].

\bibitem{websiteclavister}
Clavister.
\newblock Security in the cloud.
\newblock
  \url{http://www.clavister.com/documents/resources/white-papers/clavister-whp-security-in-the-cloud-gb.pdf}.
\newblock [Accessed July 5th, 2012].

\bibitem{websitesarbanesox}
{Congress of the United States}.
\newblock {Sarbanes-Oxley Act}.
\newblock \url{http://thomas.loc.gov}, 2002.
\newblock [Accessed July 5th, 2012].

\bibitem{dykstraacquiring}
J.~Dykstra and A.~Sherman.
\newblock Acquiring forensic evidence from infrastructure-as-a-service cloud
  computing: Exploring and evaluating tools, trust, and techniques.
\newblock {\em DoD Cyber Crime Conference}, January 2012.

\bibitem{fbi2008}
FBI.
\newblock Annualreport for fiscal year 2007.
\newblock 2008 Regional Computer Forensics Laboratory Program, 2008.
\newblock [Accessed July 5th, 2012].

\bibitem{websitegartnernews}
Gartner.
\newblock Worldwide cloud services market to surpass \$68 billion in 2010.
\newblock \url{http://www.gartner.com/it/page.jsp?id=1389313}, 2010.
\newblock [Accessed July 5th, 2012].

\bibitem{goodrich2002efficient}
M.~Goodrich, R.~Tamassia, and J.~Hasi{\'c}.
\newblock An efficient dynamic and distributed cryptographic accumulator.
\newblock {\em Information Security}, pages 372--388, 2002.

\bibitem{grisposcalm}
G.~Grispos, T.~Storer, and W.~Glisson.
\newblock Calm before the storm: The challenges of cloud computing in digital
  forensics.
\newblock {\em International Journal of Digital Crime and Forensics (IJDCF)},
  2012.

\bibitem{input2009}
INPUT.
\newblock Evolution of the cloud: The future of cloud computing in government.
\newblock
  \url{http://iq.govwin.com/corp/library/detail.cfm?ItemID=8448&cmp=OTC-cloudcomputingma042009},
  2009.
\newblock [Accessed July 5th, 2012].

\bibitem{kent2006guide}
K.~Kent, S.~Chevalier, T.~Grance, and H.~Dang.
\newblock Guide to integrating forensic techniques into incident response.
\newblock {\em NIST Special Publication}, pages 800--86, 2006.

\bibitem{khajeh2010cloud}
A.~Khajeh-Hosseini, D.~Greenwood, and I.~Sommerville.
\newblock Cloud migration: A case study of migrating an enterprise it system to
  iaas.
\newblock In {\em proceedings of the 3rd International Conference on Cloud
  Computing (CLOUD)}, pages 450--457. IEEE, 2010.

\bibitem{lunn2000computer}
D.~Lunn.
\newblock Computer forensics--an overview.
\newblock {\em SANS Institute}, 2002, 2000.

\bibitem{Ma:2009:NAS:1502777.1502779}
D.~Ma and G.~Tsudik.
\newblock A new approach to secure logging.
\newblock {\em Trans. Storage}, 5(1):2:1--2:21, Mar. 2009.

\bibitem{websitemarketresearchmedia}
{Market Research Media}.
\newblock Global cloud computing market forecast 2015-2020.
\newblock
  \url{http://www.marketresearchmedia.com/2012/01/08/global-cloud-computing-market/}.
\newblock [Accessed July 5th, 2012].

\bibitem{marty2011cloud}
R.~Marty.
\newblock Cloud application logging for forensics.
\newblock In {\em In proceedings of the 2011 ACM Symposium on Applied
  Computing}, pages 178--184. ACM, 2011.

\bibitem{reilly2011cloud}
D.~Reilly, C.~Wren, and T.~Berry.
\newblock Cloud computing: Pros and cons for computer forensic investigations.
\newblock 2011.

\bibitem{ristenpart2009hey}
T.~Ristenpart, E.~Tromer, H.~Shacham, and S.~Savage.
\newblock Hey, you, get off of my cloud: exploring information leakage in
  third-party compute clouds.
\newblock In {\em Proceedings of the 16th ACM conference on Computer and
  communications security}, pages 199--212. ACM, 2009.

\bibitem{robbins2008explanation}
J.~Robbins.
\newblock An explanation of computer forensics.
\newblock {\em National Forensics Center}, 774:10--143, 2008.

\bibitem{ruan2011cloud}
K.~Ruan, J.~Carthy, T.~Kechadi, and M.~Crosbie.
\newblock Cloud forensics: An overview.
\newblock In {\em proceedings of the 7th IFIP International Conference on
  Digital Forensics}, 2011.

\bibitem{Schneier:1999:SAL:317087.317089}
B.~Schneier and J.~Kelsey.
\newblock Secure audit logs to support computer forensics.
\newblock {\em ACM Trans. Inf. Syst. Secur.}, 2(2):159--176, May 1999.

\bibitem{taylor2010digital}
M.~Taylor, J.~Haggerty, D.~Gresty, and R.~Hegarty.
\newblock Digital evidence in cloud computing systems.
\newblock {\em Computer Law \& Security Review}, 26(3):304--308, 2010.

\bibitem{websitetikalopenstack}
Tikal.
\newblock {Experimenting with OpenStack Essex on Ubuntu 12.04 LTS under
  VirtualBox}.
\newblock \url{http://bit.ly/LFsVUY}, 2012.
\newblock [Accessed November 30th, 2012].

\bibitem{vacca2005computer}
J.~Vacca.
\newblock {\em Computer forensics: computer crime scene investigation},
  volume~1.
\newblock Delmar Thomson Learning, 2005.

\bibitem{wiles2007best}
J.~Wiles, K.~Cardwell, and A.~Reyes.
\newblock {\em The best damn cybercrime and digital forensics book period}.
\newblock Syngress Media Inc, 2007.

\bibitem{Yavuz2009baf}
A.~Yavuz and P.~Ning.
\newblock Baf: An efficient publicly verifiable secure audit logging scheme for
  distributed systems.
\newblock In {\em Computer Security Applications Conference, 2009. ACSAC '09.
  Annual}, pages 219 --228, dec. 2009.

\bibitem{zafarullah2011digital}
Z.~Zafarullah, F.~Anwar, and Z.~Anwar.
\newblock Digital forensics for eucalyptus.
\newblock In {\em Frontiers of Information Technology (FIT)}, pages 110--116.
  IEEE, 2011.

\end{thebibliography}

\end{document}